\documentclass[twocolumn,prb]{revtex4}

\usepackage{graphicx,amsmath}

\begin{document}

\title{Superconductor-insulator transition in Coulomb disorder}

\author{B. I. Shklovskii}

\affiliation{Theoretical Physics Institute, University of
Minnesota, Minneapolis, Minnesota 55455}

\date{\today}

\begin{abstract}

Superconductor-insulator transition driven by the decreasing
concentration of electrons $n$ is studied in the case of the
disorder potential created by randomly positioned charged
impurities. Electrons and Cooper pairs (formed by an non-Coulomb
attraction) nonlinearly screen the random potential of impurities.
Both electrons and Cooper pairs can be delocalized or localized in
the resulting self-consistent potential. The border separating the
superconductor and insulator phases in the plane of the
concentration of electrons and the length of the Cooper pair is
found. For a strong disorder the central segment of this border
follows the BEC-BCS crossover line defined for a clean sample.

\end{abstract}
\maketitle

Superconductor-insulator (SI) transition remains a challenging and
controversial subject for more than two
decades~\cite{Fisher,Finkelshtein,Hebard,Goldman,Nandini,Mason,Meir,Baturina,Kapitulnik}.
One way how SI transition with the decreasing concentration of
electrons $n$ can be envisioned is localization of Cooper pairs in
a random potential~\cite{Fisher}. This approach is good when
Cooper pairs weakly overlap and can be considered as repelling
each other point-like bosons, which at small external disorder
experience Bose-Einstein condensation (BEC). At large disorder the
condensate is fragmented and becomes a Bose insulator. If $\xi$ is
the size of the Cooper pair created by an (unspecified here)
attractive interaction between electrons the condition of the weak
overlap between pairs can be written as $n\xi^3 \ll 1$.

In the opposite case, $n\xi^3 \gg 1$ one may better think about
separate electrons, which can be localized or delocalized by
disorder. In delocalized state, the same attraction leads to the
Bardeen-Cooper-Schrieffer (BCS) superconductivity. Thus, at small
enough temperatures the metal-insulator transition with the
decreasing concentration of electrons leads to the SI transition
in the fermion picture as well. In this case, however, SI
transition leads into a Fermi insulator.

The goal of this paper is to study the zero temperature SI
transition phase diagram in the plane ($\xi$,$n$). We suggest a
model in which transition to Bose and Fermi insulator happens at
strongly different concentrations. These two segments of the SI
border line are connected by the long intermediate segment, where
$n\xi^3 = 1$. Here and everywhere in this paper we use the scaling
approach in the large parameter introduced below and drop all
numerical coefficients.

The line $n\xi^3 = 1$ is called BEC-BCS crossover line and in a
clean sample it is not related to any phase
transitions~\cite{Leggett,Nozi}. We, however, show that in our
model a long segment of BEC-BCS crossover line plays the role of
SI border line.

One can not meaningfully describe localization of bosons without
including their repulsive interaction. Therefore, we include the
Coulomb repulsion of Cooper pairs in our theory. In other words,
in the boson limit Cooper pairs screen random potential.
Similarly, in the BCS limit of almost free electrons, the random
potential is screened by electrons. We assume that the disorder
itself is of a the Coulomb origin. Namely, we assume that in a
three dimensional sample there are randomly distributed positive
donors with the concentration $N_D$ and negative acceptors with
the concentration $N_A$, while the concentration of electrons is
much smaller than both of them, $n = N_D - N_A \ll N$, where $N =
N_D + N_A$ is the total concentration of charged impurities. (This
model is similar to a heavily doped strongly compensated
semiconductor~\cite{SE,SEbook}). The Coulomb nature of both
disorder potential and interaction between electrons and compact
pairs makes understanding of screening very easy.

\begin{figure}[b]
\centerline{\includegraphics[width=1.8in]{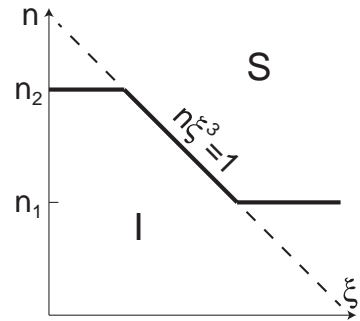}}
\caption{ \label{Fig:Diagram} The phase diagram of the
superconductor-insulator transition. On the horizontal axis we
plot the length of the pair $\xi$ (in units $N^{-1/3}$) and on
vertical line we plot the electron concentration $n$ (in units
$N$) both are in logarithmic scale. S and I stand for the
superconductor and insulator. The full line is the border between
the superconductor and insulator phases. Its sharp corners are
results of schematic nature of this drawing and actually are
rounded. The dashed line corresponds to $n\xi^{3} = 1$. The lower
segments of this line is crossover between Fermi and Bose
insulators, the upper one is discussed in the text.}
\end{figure}

It is known that on the fermion side of the diagram ($n\xi^3 \gg
1$) even at large concentration of electrons $n\sim N$ all
electronic states are localized if $Na^3 \ll 1$, where $a =
\kappa\hbar^{2}/me^2$ is the effective Bohr radius in the sample,
$\kappa$ is its dielectric constant and $m$ is the effective mass.
Thus we assume that the sample is heavily doped, $Na^3 \gg 1$.
This of course requires a large $\kappa$ or a small effective mass
$m$ leading to large $a \gg a_0$, where $a_0$ is the lattice
constant. In this model the disorder is characterized by the
single parameter $Na^3 \gg 1$. This the parameter which is used
throughout the paper for scaling estimates. Most of the
theoretical models of the SI transition studied in literature do
not have such a parameter and, therefore, have to rely on computer
simulations.

An example, where this model may be applicable, is the narrow gap
semiconductor PbTe which becomes a superconductor~\cite{Geballe}
with $T_c \sim 1.5$K, when doping by $\sim 1.5\%$ of Tl
(Pb$_{1-x}$Tl$_{x}$Te with $x \sim 0.015$) provides concentrations
of holes up to $n \sim 10^{20}~$cm$^{-3}$. The dielectric constant
of PbTe is very large~\cite{Kanai}, $\kappa > 400$. As a result
the Bohr radius $a$ of the Tl acceptor state is so large that no
freeze out is observed, i. e. for any studied doping condition of
heavy doping $Na^3 \gg 1$ (where $N$ is the concentration of Tl)
is fulfilled. It is known that superconductivity of PbTe can be
induced only by Tl doping. It is believed~\cite{Geballe,Schmalian}
that Tl is essential for attraction between electrons which
happens via quantum valence fluctuations of Tl impurities. One,
therefore, can imagine using Tl concentration to tune attraction
and, therefore, the pair length $\xi$, while the concentration of
holes $n$ can be tuned by doping with additional donors, which can
compensate Tl and make $n$ small enough for SI transition.

Our results are summarized in Fig.~\ref{Fig:Diagram}. The full
line shows the SI border. It consists of the two horizontal lines
connected by the central segment of the BEC-BCS crossover line $n
\xi^{3} = 1$. The lower horizontal line
\begin{equation}
n = n_{1}(N) = \frac{N}{(Na^3)^{1/3}}~, \label{n1}
\end{equation}
is the border between the superconductor and the Fermi insulator
phases. There is no dependence on $\xi$ here, because electrons
are only weakly bound and screen the random potential of charged
impurities like free ones. The upper horizontal line
\begin{equation}
n = n_{2}(N) = \frac{N}{(Na^3)^{1/5}} \label{n2}
\end{equation}
is the border between the superconductor and Bose insulator
phases. Again $\xi$ is irrelevant, because at such small $\xi$ the
pairs only weakly overlap and can be considered as point-like
bosons. For a given $Na^3 \gg 1$ we have got $n_{2}(N) \gg
n_{1}(N)$, because bosons have less kinetic energy and one needs
more bosons for delocalization (see derivations of $n_{1}(N)$ and
$n_{2}(N)$ below).

As we mentioned above between the two horizontal lines the SI
border follows the BEC-BCS crossover line $n\xi^{3} = 1$.
Remarkably this part of the border is disorder independent. It
exists, however, only for a heavy enough doping, when $Na^3 \gg
1$. At $Na^3 \sim 1$ we would get $n_{2}(N) \sim n_{1}(N)$ and the
intermediate range, where the border follows $n\xi^{3} = 1$,
shrinks to zero. We see again how the whole scaling picture rests
on the existence of the large parameter $Na^3 \gg 1$.

The origin of the disorder independent intermediate part of the SI
border can be understood in the following way. Suppose, we cross
this border from below, where we deal with compact Cooper pairs,
which are localized by disorder because $n < n_{2}(N)$. Thus,
crossing leads to almost free electrons with the concentration $n$
larger than $n_{1}(N)$. Electrons with such a concentration are
delocalized. Thus, the crossing of the line $n\xi^{3} = 1$ leads
from the Bose insulator to the superconductor phase. We see that
in the presence of a strong disorder the BEC-BCS crossover line
gets the new meaning.

Let us make the argument for the intermediate segment of the SI
border more formal. We know that the asymptotic segments $n = n_1$
and $n = n_2$ terminate at the line $n\xi^3 = 1$ (the two corners
of the border line in Fig.~\ref{Fig:Diagram}). It is natural to
assume that between the two termination points there is only one
intermediate physical regime or, in other words,  these two points
are connected by a power law $n(\xi)$ dependence. The only power
law connecting these points, is, of course, $n\xi^3 = 1$, i.e. the
line of BCS-BEC crossover. Thus, the SI transition border should
stick to this line at $n_2 \gg n \gg n_1$.

Let us now derive the concentrations $n_{1}(N)$ and $n_{2}(N)$.
Actually, the concentration $n_{1}(N)$ was derived in
Ref.~\cite{SE} as the metal-insulator transition in a heavily
doped strongly compensated semiconductor. It was shown also to be
in a good agreement with the experimental data for compensated
semiconductors~\cite{SEbook}. Below we repeat the derivation of
$n_{1}(N)$, because it is a necessary step for our derivation of
$n_{2}(N)$.

Let us divide the sample in cubes with the edge length equal $R$.
Due to spacial fluctuations of the concentrations of donors and
acceptor each cube has a random sign charge with the absolute
value of the order of $e(NR^3)^{1/2}$. Such randomly alternating
charges create the random potential energy relief of the amplitude
\begin{equation}
eV(R)\sim \frac{e^{2} (NR^3)^{1/2}}{\kappa R} =  \frac{e^{2}
(NR)^{1/2}}{\kappa}~. \label{VR}
\end{equation}
This energy diverges at large $R$, so that screening even by a
small concentration of electrons $n$ is crucial. To discuss this
screening let us estimate the characteristic fluctuating density
$\delta N(R)$ of charge for fluctuations with the characteristic
scale $R$. Clearly
\begin{equation}
\delta N(R) = \frac{(NR^3)^{1/2}}{R^3}=
\left(\frac{N}{R^{3}}\right)^{1/2}~. \label{DN}
\end{equation}
The concentration $n$ of electrons can be redistributed between
wells and hills of the random potential. This redistribution
screens all the scales $R$ for which $\delta N(R) = (N/R^3)^{1/2}
\leq n$ or, in other words, for $R \geq R_s$, where
\begin{equation}
R_s = \left(\frac{N}{n^{2}}\right)^{1/3} \label{RS}
\end{equation}
is the nonlinear screening radius~\cite{SE,SEbook}. All the scales
with $R < R_s$ remain unscreen, because even when all electrons
are transferred from all the hills of the potential energy to all
its wells they are not able to level off the charge density of
such fluctuations. Since $V(R) \propto R^{1/2}$ and thus grows
with $R$, among remaining scales the most important contribution
to the random potential is given by $R=R_s$. Thus, the amplitude
of the nonlinearly screened random potential energy is
\begin{equation}
eV(R_s)=\frac{e^{2}}{\kappa}\frac{N^{2/3}}{n^{1/3}}~. \label{V}
\end{equation}
So far we dealt only with the electrostatic energy of electrons
and neglected their kinetic energy. We are talking about the limit
of zero temperature, so that all the kinetic energy is of the
quantum origin. Now we should find conditions when the quantum
kinetic energy is small enough so that described above regime of
localized electrons is valid. Clearly the potential energy
Eq.~(\ref{V}) is able to localize electrons with concentration $n$
if it is larger than the Fermi energy of electrons
$\epsilon_{F}(n) = \hbar^2 n^{2/3}/2m$ in its wells. In the
opposite case $\epsilon_{F}(n) \gg eV(R_s)$ the Fermi sea covers
the typical maxima of the potential energy and the semiconductor
behaves as a good metal. Equating $eV(R_s)$ and $\epsilon_{F}(n)$
we arrive at the critical concentration $n_{1}(N)$ of the SI
transition given by Eq.~(\ref{n1})~\cite{SE,SEbook}.

In the metallic phase electron screening becomes linear, the
screening radius is given by the standard Thomas-Fermi expression
$R_{TF}= a/(na^3)^{1/6}$ and the amplitude of the screened
potential relief is equal $eV(R_{TF}) =
e^{2}(NR_{TF})^{1/2}/\kappa$. These quantities match
Eqs.~(\ref{RS}) and (\ref{V}) at $n=n_1$.

Let us now switch to calculation of the critical concentration
$n_{2}(N)$ of electrons in charge $2e$ bosons, corresponding to
the SI transition to the Bose insulator. For the insulating phase
of localized composite bosons we can exactly repeat the above
calculation of the nonlinear screening radius $R_s$ and the random
potential energy created by screened charged impurities (\ref{V}).

The difference between the gas of composite bosons and that of
fermions becomes important only at the last step of calculation of
the critical concentration, where one has to consider limitations
caused by the quantum kinetic energy. Many composite bosons can
occupy one localized level of each potential well. Therefore,
condition of delocalization of composite bosons is much stronger
than the condition $eV(R_s) < \epsilon_{F}(n)$ used for fermions.
Namely, for delocalization of composite bosons we should require
that a typical well of the random potential does not have a level,
or $eV(R_s) < \hbar^{2}/mR_{s}^2$. Solving the equation
\begin{equation}
eV(R_s) = \frac{\hbar^{2}}{mR_{s}^2}~, \label{BE}
\end{equation}
for $n$ and using Eqs. (\ref{V}) and (\ref{RS}) we get the
critical concentration of the SI transition Eq.~(\ref{n2}). This
derivation clearly shows why $n_{2}(N) \gg n_{1}(N)$. The
potential energy amplitude $eV(R_s)$ according to Eq.~(\ref{V})
decreases with increasing $n$. In order to achieve delocalization
in the composite boson case, we had to make $eV(R_s)$ smaller than
in the fermion case and this requires the larger concentration
$n_{2}(N) \gg n_{1}(N)$.

One can arrive at the critical concentration $n_{2}$ from the
large $n$ superconductor phase of delocalized Cooper pairs, as
well. For this purpose we have to start from screening in the case
when a composite boson can not be localized inside the well of the
size $R$ and in the first approximation its wave function $\psi_0$
is the plane wave with a very small wave-vector. In the random
potential energy with characteristic length $R$ and the amplitude
$2eV(R) \ll \hbar^{2}/mR^2$ this wave function is slightly
modulated with the small amplitude $\delta \psi = \psi_{0}
[eV(R)/(\hbar^{2}/mR^2)]$. For the amplitude of the fluctuating
density of charge of electrons with the scale $R$ this gives
\begin{equation}
\delta n(R) = n \frac {eV(R)}{(\hbar^{2}/mR^2)}~. \label{Dn}
\end{equation}
All scales $R$ of charge fluctuations for which $\delta n(R) \geq
\delta N(R)$ can be screened. Therefore, the equation $\delta n(R)
= \delta N(R)$ defines the radius $R_d$ of linear screening by
delocalized composite bosons. Using Eqs.~(\ref{Dn}) and (\ref{DN})
we get
\begin{equation}
R_d = \left(\frac{a}{n}\right)^{1/4}~. \label{Rd}
\end{equation}
(We could not find this result in the literature.) The meaning of
this screening radius is that the Coulomb potential of all
fluctuations of charge with $R > R_d$ are screened by small
changes of wave functions, while all the fluctuation scales with
$R < R_d$ are unscreened. Using Eq.~(\ref{VR}) it is easy to see
that at $n \gg n_2$ the remaining amplitude of potential energy
fluctuations $eV(R_d) \ll \hbar^{2}/mR_{d}^{2}$, so that we indeed
deal with delocalized composite bosons. Note that as one should
expect the linear screening radius $R_d$ of the delocalized phase
matches the nonlinear screening radius $R_s$ of the localized one
at $n = n_2$. Of course, simultaneously $V(R_d)$ matches $V(R_s)$.

One can also verify that on the SI border line inequality $R_s \gg
\xi$ which we need to consider screening by composite bosons as
point like objects fails only at $\xi \sim N^{-1/3} (Na^3)^{2/9}$.
This points resides to the right of lower-right corner of SI
border line of Fig.~\ref{Fig:Diagram}. This justifies screening
calculations presented above.


So far we were concerned with the shape of the full line
separating the insulator and superconductor phases on
Fig.~\ref{Fig:Diagram}. One may ask whether the two segments of
the dashed BEC-BCS crossover line $n\xi^{3} = 1$ abandoned by the
full line make sense in the insulator phase (the right low corner
of the diagram) and in the superconductor phase (the left upper
corner). The answer is that these segments of BEC-BCS crossover
line imly important crossovers within both phases.

The dashed line in the insulator phase signals the change of a
hopping excitation with increasing $n$ from a double charged
composite boson to an electron. On both sides of the dashed line
the low temperature variable range hopping conductivity obeys the
Efros-Shklovskii law~\cite{CG}
\begin{equation}
\sigma =
\sigma_{0}\exp\left[-\left(\frac{T_0}{T}\right)^{1/2}\right]~,
\label{S}
\end{equation}
but with two different characteristic temperatures $T_0$. For the
large $n$ phase of localized individual electrons (Fermi
insulator)
\begin{equation}
T_0= \frac{Ce^2}{\kappa l}~, \label{TO}
\end{equation}
where $l$ is the electron localization length and $C\sim
2.7$~\cite{SEbook}. For the low $n$ phase of localized Cooper
pairs (Bose insulator) the double charge $2e$ replaces $e$ in
Eq.~(\ref{TO}). The localization length of a Cooper pair is also
smaller than $l$. Thus, the characteristic temperature $T_0$
substantially decreases with increasing $n$ or $\xi$ at the
BEC-BCS crossover. This leads to the steep increase of the hopping
conductivity at the dashed line. This phenomenon is similar to the
reduction of the characteristic temperature $T_0$ by the magnetic
field in a granular superconductor discussed recently in
Ref.~\cite{Lopatin}.

On the other hand, the dashed line segment in the left upper
corner of our phase diagram signals a change in the concentration
dependence of the superconductor gap and the critical temperature
of the superconductor-metal transition. At this line both
quantities start decreasing sharply with the growing
$n$~\cite{Nozi}.

Above we considered $n$ and $\xi$ as the two independent
variables. Strictly speaking this is correct only for $n \xi^{3}
\leq 1$. Therefore, small $\xi$ (compact bosons) and intermediate
segments of the SI border line are found correctly (in the scaling
sense). The large $\xi$ segment of our border is above the BEC-BCS
crossover line $n \xi^{3} = 1$ and, therefore, $\xi$ may be a
function of $n$. (In BCS theory $\xi$ grows with $n$). This,
however, can not affect validity of our result Eq.~(\ref{n1}),
because the large $\xi$ segment of the border line is not
sensitive to $\xi$ at all. This justifies our results for the
whole SI border.

Until now we dealt with systems where both impurities and
electrons reside in three dimensions (we can call it 3D-3D case).
We do not know a 3D-3D system, where $n$ can be made tunable in
one sample. More promising can be the 3D-2D case, where impurities
residing in three dimensional space (with the three-dimensional
concentration $N$) surround a two-dimensional system of electrons
(with the two-dimensional concentration $n$) and affect it by the
Coulomb random potential. This can be a very thin superconductor
film on the top of strongly compensated insulator such as a low
quality SiO substrate. In such geometry concentration $n$ can be
also regulated by a gate\cite{Parendo}.

For 3D-2D case one can follow the logic of the calculations above
and arrive at a similar phase diagram in the plane ($\xi$,$n$).
Let us first show the results for this case and then outline the
calculation. For the critical concentrations where localization of
electrons and composite bosons takes place $n_1$ and $n_2$ we
arrived at
\begin{equation}
n_{1} = \frac{N^{2/3}}{(Na^3)^{1/6}}~,~~~~~~~~n_{2} =
\frac{N^{2/3}}{(Na^3)^{1/15}}~, \label{n32D}
\end{equation}
where $a$ is the Bohr radius of two dimensional electrons. The two
horizontal segments of the border line $n = n_1$ and $n=n_2$ are
connected by the BCS-BEC crossover line $n\xi^{2}=1$, so that the
phase diagram in the plane ($n$, $\xi$) looks similar to
Fig.~\ref{Fig:Diagram}.

The main steps of this calculation are as follows. We again start
from from cutting three-dimensional space filled by impurities in
cubes with the edge length equal $R$ in such a way that on each
side of the plane  of the two-dimensional electron gas the first
layer of cubes touches the plane. These two first layers create
the "projected to the plane" two-dimensional fluctuating charge
density of impurities $(NR^3)^{1/2}/R^{2}$. Equating this density
to $n$ we find the nonlinear screening radius $R_s = N/n^{2}$.
Then we calculate the amplitude of the nonlinearly screened random
potential energy substituting $R = R_s$ into Eq.~(\ref{VR}). This
gives
\begin{equation}
eV(R_s) = \frac{e^{2}}{\kappa}\frac{N}{n}~. \label{VR2}
\end{equation}
Equating this energy to the Fermi energy $\hbar^{2}n/m$ yields
$n_1$, while equating it to $\hbar^{2}/mR_{s}^{2}$ gives $n_{2}$.
In derivation of Eq.~(\ref{VR2}) we followed Ref.~\cite{QHESE}
where similar problem was studied in order to calculate
thermodynamic density of states at a small filling factor of a
topmost Landau level in quantum Hall effect regime in a MOSFET or
a GaAlAs heterostructure without a spacer.

In conclusion, we suggested a model with a strong Coulomb disorder
and strong Coulomb interaction, which lets us construct the
scaling phase diagram of the SI transition in the plane of the
concentration of electrons $n$ and the characteristic length of
the Cooper pair $\xi$ both in 3D and 2D systems. The SI border
consists of three segments. Two asymptotic segments for a large
and small lengths of Cooper pairs describe transitions in the
limits of almost free fermions and of compact bosons. They
correspond to the two different concentrations of electrons $n_1$
and $n_2$. In the strong disorder the threshold concentration for
compact Cooper pairs, $n_2$, is much larger than that for almost
free electrons, $n_1$. As a result for a strong disorder the two
asymptotic segments of the border line are connected by the third
long segment, which follows the line of BEC-BCS crossover
conventionally defined for a clean sample~\cite{Leggett,Nozi}.
Thus, our model demonstrates that the strong disorder can bring
about an important implication for the BEC-BCS crossover line: it
becomes the border line of the SI transition.

I am grateful to M. Fogler, Tao Hu, P. Lee, S. Sachdev, J.
Schmalian, D. Sheehy, B. Spivak, and P. Wolfle for useful
discussions. I acknowledge hospitality of the Aspen Center for
Physics, where the first draft of this paper was written.


\end{document}